\pdfoutput=1

\makeatletter
\providecommand{\doi}[1]{%
	\begingroup
	\let\bibinfo\@secondoftwo
	\urlstyle{rm}%
	\href{http://dx.doi.org/#1}{%
		doi:\discretionary{}{}{}%
		\nolinkurl{#1}%
	}%
	\endgroup
}
\makeatother

\documentclass[final,5p,times,twocolumn]{elsarticle}

\usepackage{amssymb}
\usepackage{bm}
\usepackage{graphicx}
\usepackage{booktabs}

\usepackage{multirow} 
\usepackage{rotating}
\usepackage{natbib}
\usepackage{hyperref}

\usepackage{lineno}
\usepackage{physics}
\usepackage{hyperref}
\journal{Journal of Electron Spectroscopy and Related Phenomena}
\begin{document}
\begin{frontmatter}

\title{Quantitative electron spectroscopy of $^{125}$I over an extended energy range }

\author[EME,SA]{M. Alotiby}
\author[ANSTO]{I. Greguric}
\author[NUCLEAR]{T. Kib\'{e}di} 
\author[NUCLEAR]{B. Tee}
\author[EME]{M. Vos} \ead{maarten.vos@anu.edu.au}

\address[EME]{Electronic Materials Engineering, Research School of Physics and Engineering, Australian National University, Canberra, ACT Australia}
\address[SA]{King Abdulaziz City for Science and Technology, Riyadh, Saudi Arabia}
\address[ANSTO]{ Australian Nuclear Science and Technology Organisation, Lucas Heights, NSW, Australia}
\address[NUCLEAR]{Nuclear Physics, Research School of Physics and Engineering, Australian National University, Canberra, ACT, Australia}
\vspace{10pt}
\date{}

\begin{abstract}
Auger electrons emitted after nuclear decay have potential application in targeted cancer therapy. For this purpose it is important to know the Auger electron yield per nuclear decay. In this work we measure the ratio of the number of  conversion electrons (emitted as part of the nuclear decay process) and the number of  Auger electrons (emitted as part of the atomic relaxation process after the nuclear decay) for the case of $^{125}$I. Results are compared with Monte-Carlo type simulations of the relaxation cascade using the BrIccEmis code. With the appropriate choice of parameters this program describes the observed spectra quite well over the whole energy range.
\end{abstract}

%
%
%
%
%
\begin{keyword}
	
	
Auger electron spectroscopy; iodine-125; conversion electrons
\end{keyword}

\end{frontmatter}

\section{Introduction}
For treatment of cancer a very localised radiation source that  only affects the body within a cancer cell is highly desirable. Low energy ($<$~1000 eV) electrons have a very short mean free path (of the order of nm) for
inelastic excitations and hence a short range. The corresponding high linear energy transfer (LET) values are attractive if
one aims to target cancer cells and minimise collateral damage to neighbouring healthy cells.  A convenient way to produce such low-energy electron radiation could be provided by isotopes that emit Auger electrons as part of their nuclear decay.  Their possible use in cancer therapy has been discussed
extensively (see e.g. \cite{Howell1992,Pomplun2016,Rezaee2017,Bavelaar2018}). 

In order to get a good understanding of  the potential of an isotope for `Auger therapy' it is thus imperative to have precise
knowledge of the number of Auger electrons emitted per nuclear decay, as well as their energy distribution.  The Auger electrons are emitted as part of the atomic  relaxation process, initiated by an inner-shell vacency produced by the nuclear decay event. Atomic relaxation is a
complex process with many possible pathways, especially for higher atomic numbers.  The problem is
most conveniently tackled using  Monte Carlo simulations based on decay rates as calculated for 
isolated atoms \citep{Pomplun1987,Nikjoo2008,Lee2012}.  Experimental verification of the results of
these  simulations is then, of course, critical for accurate medical dosimetry. 

From the  point-of-view of electron spectroscopy this is a challenging problem. Quantitative electron spectroscopy over a wide energy range is difficult.  Fortunately, emission of Auger electrons is often preceded by the emission of conversion electrons. These electrons, emitted as part of the nuclear decay process,  can have energies that are quite similar to the Auger electrons.  The intensity of the  conversion electrons is relatively well understood.  By comparing Auger and conversion electron intensities one can `calibrate' the Auger intensity per nuclear decay.  This paper describes a first attempt to measure the complete  energy-resolved intensity distribution of emitted electrons after nuclear decay for $^{125}$I absorbed on a Au surface.

\section{Background}
The two nuclear decay processes producing inner-shell vacancies are electron capture (EC) and internal conversion. In EC an atomic (mainly inner shell) electron is absorbed by the nucleus (and a neutrino is emitted). In internal conversion an inner-shell  electron absorbs  energy from the nucleus and  is ejected. This electron is called the conversion electron (CE).

\begin{figure}
	\centering
	\includegraphics[width=6cm]{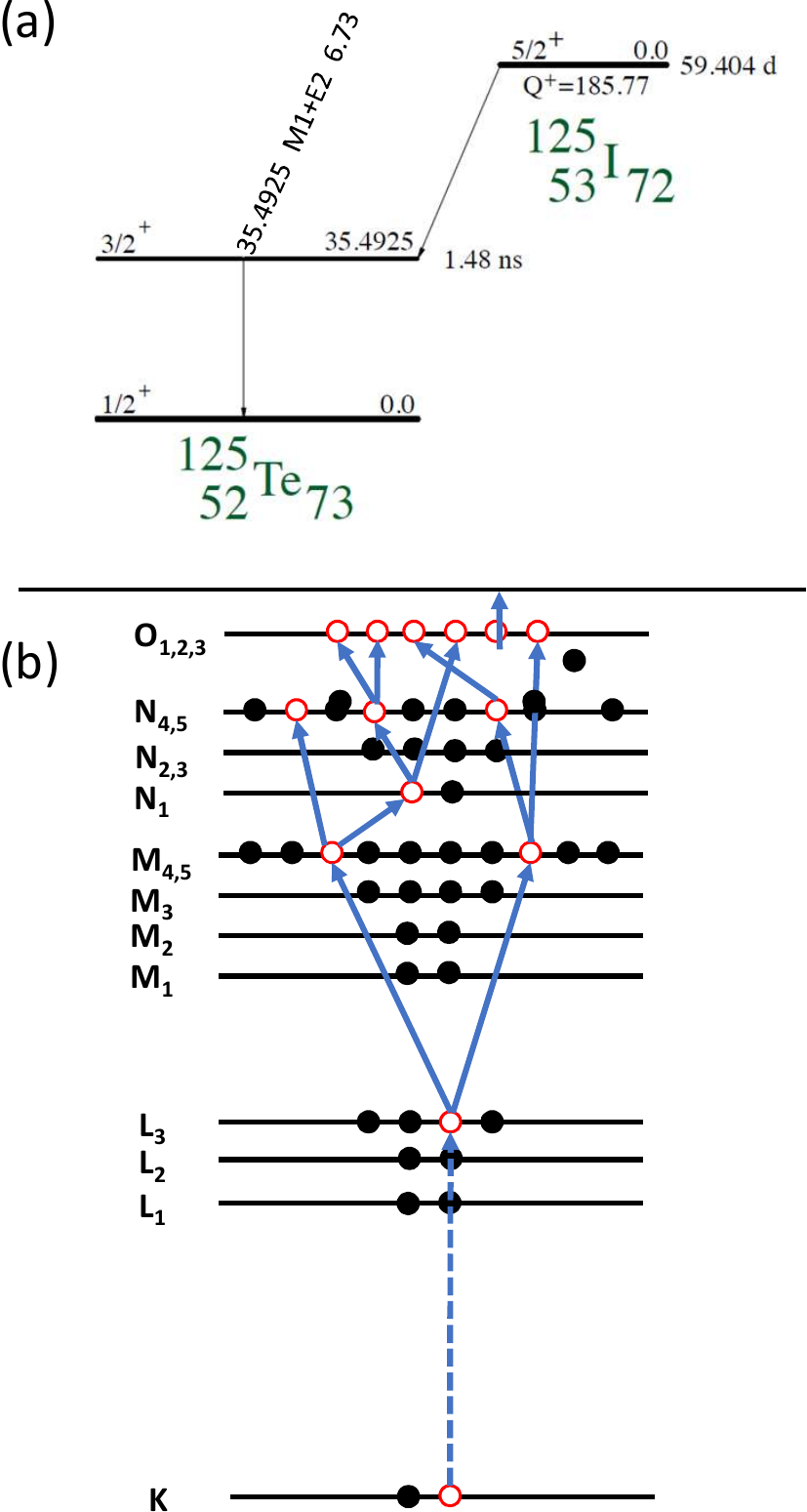}
	\caption{In (a) we show the decay scheme of $^{125}$I. An example of a relaxation cascade after core hole creation, consisting of x-ray (dashed line) and Auger electron emissions (full lines), is shown in (b)}
	\label{fig:decayschemecascade}
\end{figure}

In this study the $^{125}$I isotope was used.  Its decay scheme is shown in Fig. \ref{fig:decayschemecascade}(a). It is the proto-typical candidate for Auger therapy, e.g. it was used in the
original Monte Carlo simulations of ref. \cite{Pomplun1987}.  $^{125}$I decays with a half-life of 59.4 days via electron capture to an excited state of
$^{125}$Te. The excited nucleus decays in 93\% of the cases to its ground state by the emission of a CE.
The half-life of this excited state  is 1.48 ns, which is much longer than the time scale for  relaxation of the electronic structure of an atom to its ground state (femtoseconds). The energy of the excited state is 39.4925 keV and the resulting K CEs have an energy of $\approx 3.678$ keV, within the range of the energies of the Te LMM Auger electrons. If an L$_1$ electron is removed, the CE has an  energy of 30.553 keV, 20\% more than the KLL Auger electrons.

Core hole creation is followed by a sequence of Auger and/or X-ray emissions (see Fig. \ref{fig:decayschemecascade}(b)).  There are  many different cascade sequences possible and a Monte Carlo program BrIccEmis was developed by Lee {\it et al.} to describe this \cite{Lee2012}. 
For the decay of $^{125}$I there are two separate relaxation cascades contributing   to the Auger yield: the first one after electron capture and the second one after emission of a CE.
The combined large Auger yield makes  $^{125}$I an attractive candidate for targeted cancer therapy. The first generation of  studies of Auger and CE emission by   $^{125}$I used magnetic spectrometers, see e.g. \cite{Graham1962,Casey1969,Brabec1982}.   Recent experimental work on Auger emission after decay of  $^{125}$I, using an electrostatic analyser,
\citep{Pronschinske2015}  indicates an enhancement  of the very-low energy electron yield. A summary of  results for the KLL and LMM Auger electrons was given elsewhere \cite{Alotiby2018}.  Here we want to describe  the whole spectrum and discuss the experimental methodology  more in-depth.

\section{Experimental Details}
$^{125}$I  can be prepared as a  sub-monolayer source on a Au(111) surface which is
stable in air \citep{Huang1997} and it has been shown that the Te atoms, produced in the decay
process, are bound to this surface as well \citep{Pronschinske2016}.  Hence $^{125}$I is a 
suitable test case for the current application.

Samples with a third of a monolayer of $^{125}$I on a Au(111) surface were prepared following the
procedure described by \cite{Pronschinske2015}.  Au(111) surfaces were obtained by flame annealing
Au samples from Arrandee Metal GMBH, Germany just before the $^{125}$I deposition. A droplet containing  NaI in a NaOH solution (pH $\approx 10$, from Perkin Elmer) was put on this surface, and left to react. In this way an approximately  4~mm diameter source
 was obtained with an activity of 6.4~MBq.

\begin{figure}
	\centering
	\includegraphics[width=7.5cm]{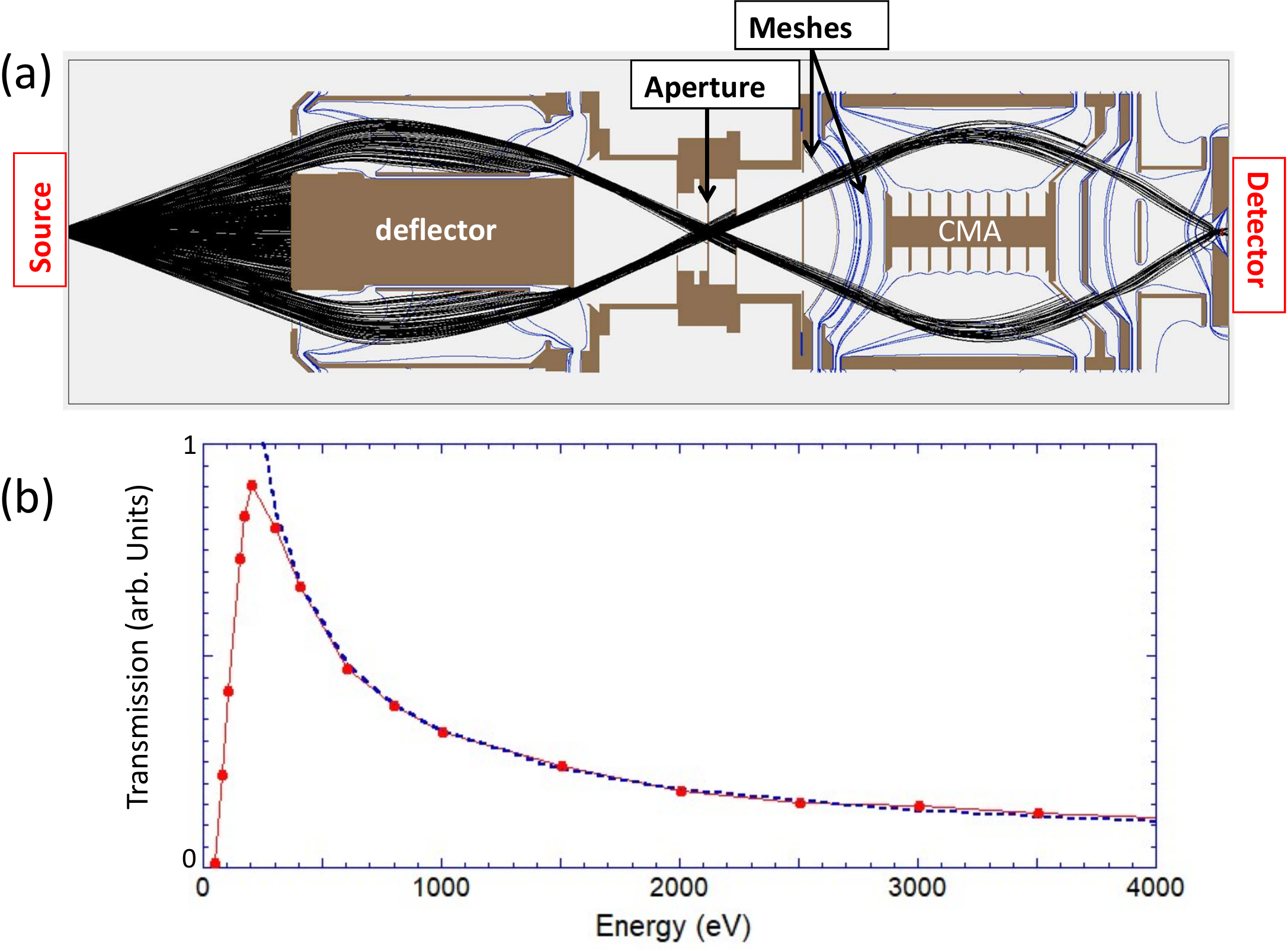}
	\caption{(a) SIMION simulation for the SuperCMA. Electrons emitted from the sample are transmitted by the deflector through an aperture. Electrons enter subsequently the floating CMA analyser stage, and electrons with the right energy are subsequently accelerated towards the detector and counted. With increasing kinetic energy  of the electrons entering the CMA the deceleration while entering the second stage causing a larger divergence  and the fraction that is counted decreases. Some of the equipotential lines are shown in blue.  (b) dots: the transmission of the analyser as simulated at 100 eV pass energy as a function of the electron energy.  Dashed line: transmission assuming a  $1/E^{0.8}$ energy dependence.  }
	\label{CMA_Simion_results} 
\end{figure}

The samples were measured  with two spectrometers.  For electrons with energies below 4 keV, a modified version of
the DESA100 SuperCMA of Staib Instruments was used, referred to in the following  as `CMA'. This is a two-stage cylindrical mirror analyser (see Fig. \ref{CMA_Simion_results}), where the 
energy of the electrons in the second stage (the `pass energy') is user-controlled to select the energy resolution. 
 The electronics was modified in order to extend the
measurement range  from 0-2.5 keV to 0-4 keV. 	To be specific, the main high-voltage power supply was
replaced by  a custom built one. The spectrometer itself was slightly modified by incorporating some
high-$Z$ metal shielding to prevent X-rays emitted by the sample interacting with the channeltron.

The second spectrometer was a locally-built spectrometer that can measure electrons with energies
between 2~keV and 40~keV \citep{Vos00,Went2005}.  It was developed originally for Electron Momentum Spectroscopy, and hence we refer to this spectrometer as `EMS'.  This spectrometer has a smaller opening angle, but it
is equipped with a two-dimensional detector making it possible to measure a range  (17\% of the pass
energy) of energies simultaneously. The sample was floated at a positive high voltage and the
hemispherical analyser was  kept near ground potential. This spectrometer was operated in two  different modes.
In the high resolution mode the pass energy was 200 eV, the  sample high-voltage was kept constant
and the analyser offset voltage was varied by up to 1 keV. Stability of the sample high-voltage was checked
using a precision voltage divider (Ross Engineering VD45) and a 7-digit volt meter, and finding it to be
better than 0.2~V. However, the absolute accuracy of the high-voltage measurement is not expected to be better than 5 V.  The energy resolution was $\approx 3$~ eV in this mode but the range of energies
that can be measured was limited to $\approx 930$~eV due to constraints on the voltage that can be
applied to the analyser.

In the low resolution mode the main 40~kV power supply was controlled by a computer using a
16-bit DAC. Measurement of the obtained voltage showed deviations up to 8 V from the requested one 
when the high voltage was varied between 2 and 35 keV (using a simple linear calibration), but when
the voltage is varied over a smaller range (1-2 keV) then this deviation was fairly constant
($\approx 1$ eV) over this range.  In this mode the pass energy was set  at 1000 eV.  The energy
resolution obtained in this way was (as will be shown later)  6 eV, but the energy range that
can be measured is not restricted and the data acquisition  rate is about 5 times higher.  If no
peaks  are expected over a particular energy range, then that energy range  was skipped,
further speeding up the measurement. Besides the main high-voltage, one of the lens voltages was
changed under computer control to ensure that the decelerating lens stack forms an image of the
source at the entrance plane of the analyser at all times.

Note that the Te LMM Auger is in an energy range that is covered by both analysers and we will exploit this overlap to obtain information from 10's of eV to 35 keV at approximately the same scale.

Additionally some electron-beam induced Auger spectra from Te films were measured for comparison with the Auger results after  $^{125}$I  decay. Thin films ($\approx$ 40 \AA\ thick) of Te were grown on thin ( $\approx$35~\AA) free-standing carbon films by thermal evaporation. Their Auger spectra were induced by a 50 nA electron beam with an energy of 29 keV  that was transmitted through these films.
 Using thin targets ensures that the background below the Auger peak does not swamp the Auger signal completely, as there are only limited possibilities of generating 3-4~keV electrons in such thin films \cite{Went2005,Went2006a}.

\begin{figure}
	\centering
	\includegraphics[width=7.5cm]{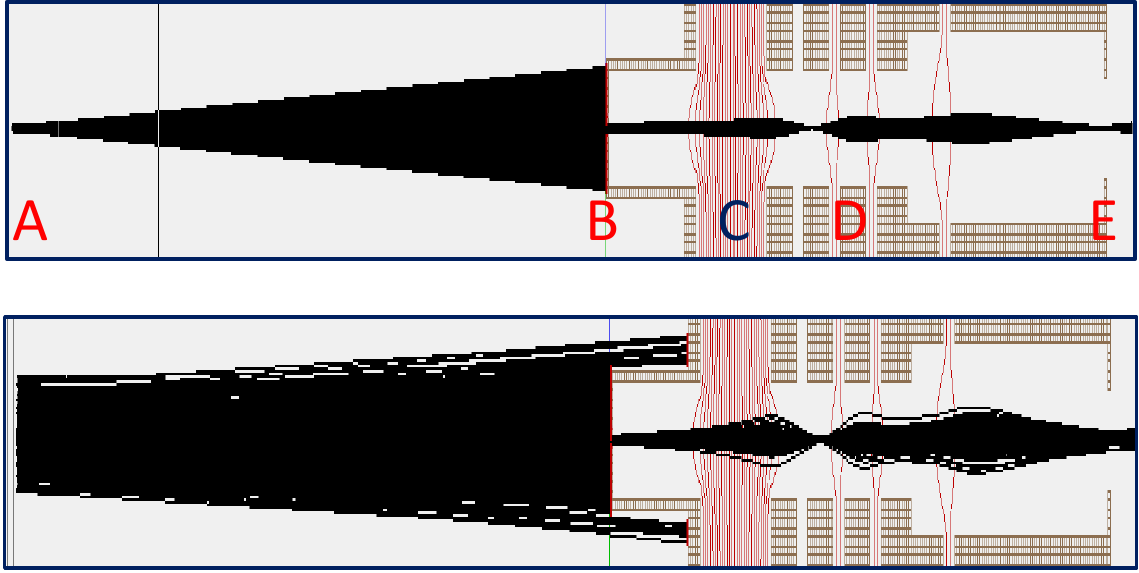}
	\caption{Examples of SIMION simulations for the high-energy spectrometer. Electrons emitted from the source (A) are restricted from entering the lens stack by a 0.5~mm wide slit (B). After the main deceleration stage (C) the electrons are focused by a set of electrostatic lenses (D) on the exit plane (E) that coincides with the entrance of the hemispherical analyser.  The vertical scale is expanded by a factor of 4, for clarity. The top panel shows the case of a 0.2~mm diameter source, as is the case when an electron beam hits the sample. The lower panel shows the case for a source size of 4~mm diameter as is the case for our $^{125}$I source.  The red lines are equipotential planes plotted at 1000 V intervals. }
	\label{Simion_results} 
\end{figure}
A main aim of this work was to establish experimentally the ratio of the intensity of CEs emitted
as part of the nuclear decay process and Auger electrons, emitted as part of the electronic
relaxation process.  There are several CE lines within the range of the high-energy spectrometer
(between 30.5 and 35.5 keV), whereas the K-CEs  at 3.679 keV can be measured with both
spectrometers. As the Auger energies  will differ from the CE energies, it is essential to understand
how the efficiency of the analysers varies with the electron energy.  

The combined transmission and
detection efficiency  of the same model CMA has been  determined experimentally by Gergeley {\it et al.} \citep{Gergeley1999} and
for energies higher than a few times its pass energy fitting of the results of  Gergeley  {\it et al.} gives an efficiency that scales like $1/E^{1.2}$, with $E$ the electron energy.  For energies
less than the pass energy the transmission  drops rapidly.  Our own  electron optics simulation, shown in Fig. \ref{CMA_Simion_results}(a), obtained using the SIMION program \cite{Dahl2000}, shows a transmission curve that resembles the result of Gergeley {\it et al.}
 but has a somewhat weaker energy dependence (proportional to  $\approx 1/E^{0.8}$) for  $E$ values significantly larger than the pass energy. 

For the high-energy spectrometer the situation is different. It uses a lens stack
that decelerates the electrons and focuses them at the exit of this lens stack, which coincides with
the entrance of an hemispherical analyser.  Examples of SIMION simulations for the lens stack are
shown in Fig. \ref{Simion_results}. In these simulations the initial kinetic energy of the electron
was 30 keV, and at the exit of the lens stack this energy was reduced to 1000~eV, the pass energy of
the  analyser. The lens stack forms an image of the source at the entrance of the
hemispherical analyser with a  magnification of $\approx$ 0.5. The simulations were done for a
0.2~mm diameter source (as is the case when an electron beam hits the sample) and a 4~mm diameter source (as is the case for our $^{125}$I sample).  In both cases
all electrons transmitted through slit (B) will enter the analyser.  The spectrometer transmission 
is thus determined solely by the width of the entrance slit and  is independent of $E$. As the
hemispherical detector transfers this image (with unity magnification) on the channel plates, the
larger spot size at the entrance of the hemisphere, when a radioactive source is used,
causes some  deterioration in energy resolution. As the energy  of  the electrons after deceleration is always the same, the detection efficiency of the channel plates does not depend on the initial kinetic energy of the detected particle.

At energies of several eV's one can obtain information about the number of electrons leaving the sample by measuring the current going towards the sample, as was demonstrated by Pronschinske {\it et al.} \citep{Pronschinske2015}. In our experiment a carbon coated, 85\% transmission mesh was placed in front of the sample. By applying a negative voltage $V_{\rm repel}$ to this mesh one can control the cut-off point of the electrons with enough momentum  along the surface normal  to leave the sample ($p_\perp^2/2m_e> -V_{\rm repel}$). If a metal plate is used, instead of a high-transparency carbon-coated mesh, then energetic Auger electrons hitting the plate can generate secondary electrons that are subsequently accelerated towards the sample by the applied voltage.  This causes the sign of the measured current to change for larger applied (negative) voltages. The use of a carbon-coated, high-transparency  mesh minimises the secondary electron emission and then such a  sign reversal was not observed.
\section{Results}
We describe the spectra starting at high energies and subsequently move to lower energies.  At high energies interpretation of the measurements is relatively  easy, and the knowledge obtained here can be used to unravel the lower energy cases where lines tend to overlap more.

\begin{center}
	\begin{figure}
		\centering
		\includegraphics[width=7cm]{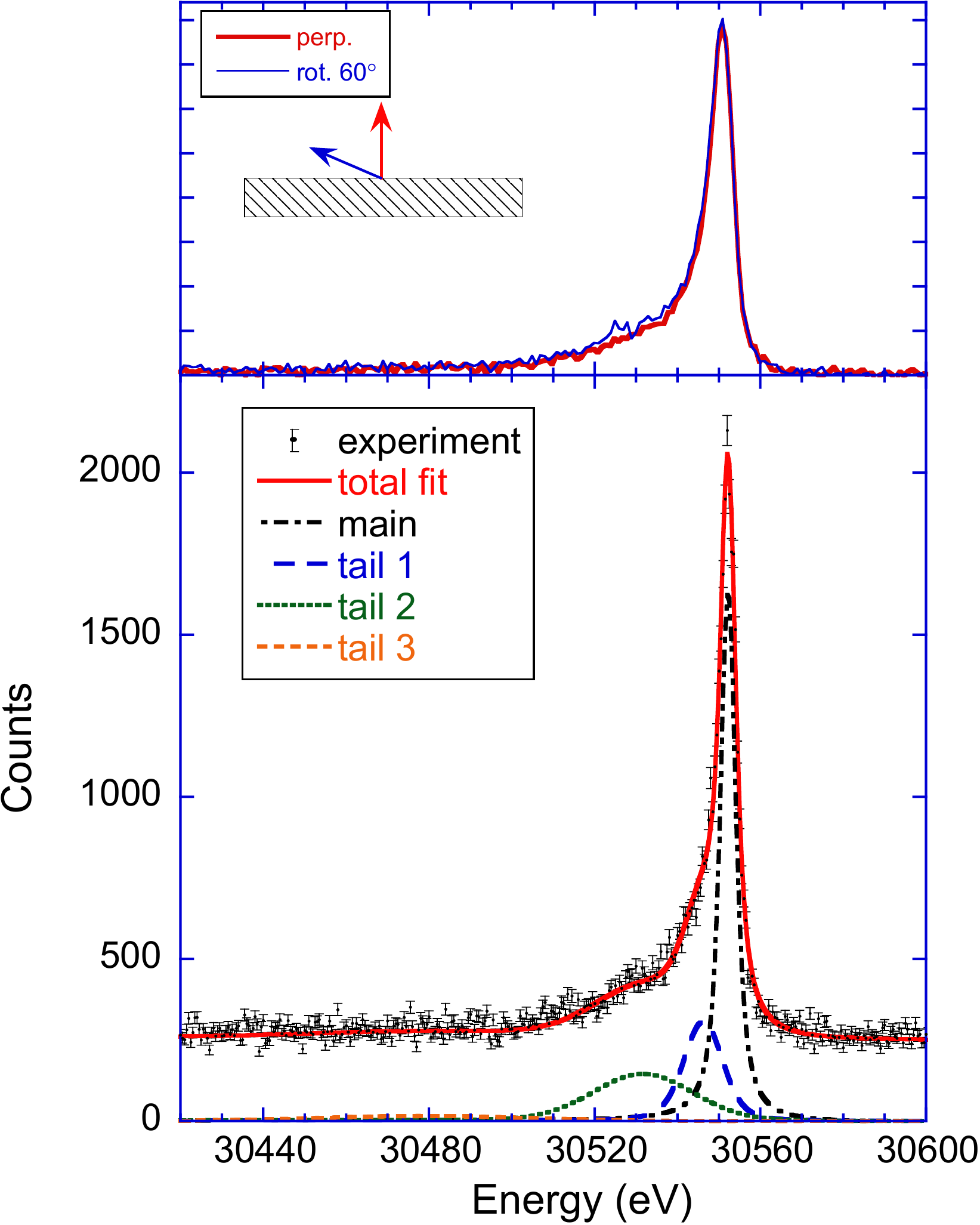}
		\caption{Lower panel: The L$_1$-CE spectrum measured at high resolution (200 eV pass energy, offset power supply scanned), compared with a fit  obtained as described in the text.  Upper pannel: the effect of rotation of the sample on the lineshape is small.}
		\label{fig:l1-high-res}
	\end{figure} 
\end{center}

\begin{table}
	\centering
	\begin{tabular}{|l|c|c|c|}
	\hline
	~      & Position (Shift)  & Rel. intensity & Width  \\
	~      & (eV)              & ~              & (eV)   \\ \hline
	Main   & 30552             & 1              & 0    \\
	Tail 1 & (-5 )            & 0.4           & 8   \\
	Tail 2 & (-18 )           & 0.5          & 24   \\
	Tail 3 & (-42)             & 0.2           & 55     \\
	\hline
\end{tabular}
	\caption{Parameters used to fit the spectrum of Fig. \ref{fig:l1-high-res} and the Gaussian width of the different components. All components were additionally broadened by an estimate of the experimental resolution (4.8 eV) and the calculated lifetime broadening of the L$_1$ level (Lorentzian of 2.2 eV FWHM).}
	\label{table:fit}
\end{table}

\begin{figure}
	\centering
	\includegraphics[width=7.5cm]{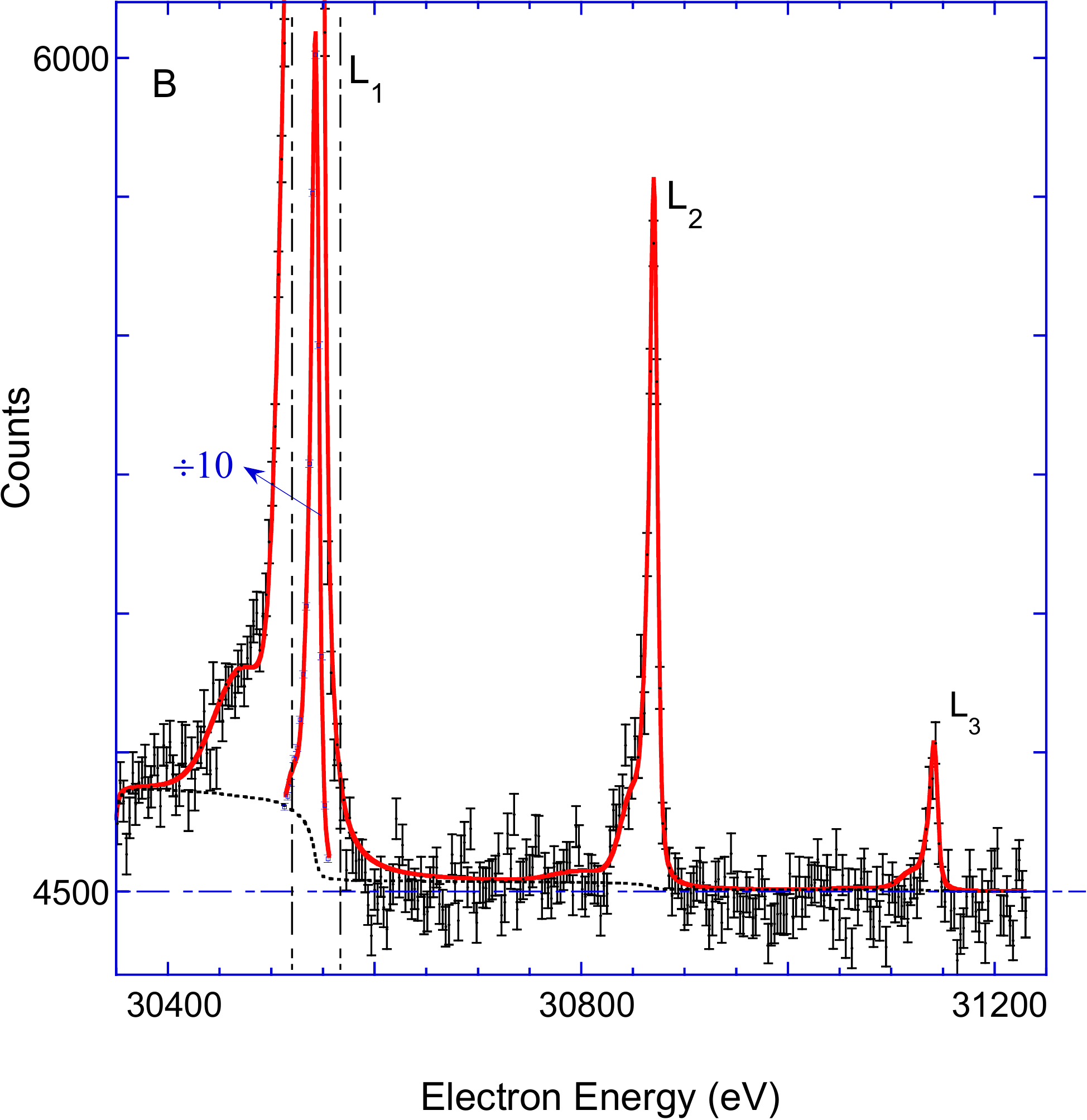}
	\caption{  The spectrum of  L$_1$, L$_2$ and L$_3$ conversion lines recorded at low resolution (main power supply scanned and 1000 eV pass energy).  The fit (red)  of the L$_1$ to L$_3$-CE spectrum uses the line shape parameters   of Table \ref{table:fit}, except that the Gaussian width is now 6 eV.}
	\label{fig:l1-l3-low-res}
\end{figure}

\subsection{High-Energy Conversion Electron Spectra}
The L$_1$ CE spectrum was used to establish the spectrometer performance and the  `shake effects' in the sample after core electron emission.  This knowledge will subsequently be used  to aid  the interpretation of the Auger spectra.
In Fig. \ref{fig:l1-high-res}, a spectrum of the L$_1$ conversion line in the
high resolution mode is shown.
A well-defined peak with a clear tail was observed at an energy of 30.552 keV, very close to the expected
value for the L$_1$ conversion line (30.553 keV). However, this level of agreement is accidental as (i) the high-voltage measurement has an estimated accuracy  of 5 eV (ii) the binding energy used is obtained relative to the Fermi level \cite{Bearden1967}, whereas the measurement is relative to the vacuum level and (iii) this value will depend on chemical shifts, not included in the calculation. 

To get a precise description of the L$_1$ CE peak shape  we employed several peak shapes but all
resulted in very similar conclusions as the one described here. 
In order to get a good fit it was required to add 2 or 3 additional Gaussian components. 
All components were convoluted with a Lorentzian representing the lifetime broadening of the L$_1$
level.  The  compilation   of the lifetime from Cambell and Papp  \cite{Campbell2001} was used for all levels  in this paper, unless stated otherwise.  Also a  small
increase in the background at the low-energy side of the peak was implemented using the Shirley approach
\cite{Shirley1972}.  The obtained  fit parameters are listed in Table \ref{table:fit}.  The Gaussian
width of the main peak is taken as the energy resolution in this operation mode ($\approx $3 eV  full width at
half maximum (FWHM)).  This is  larger than the energy resolution obtained in REELS (reflection
electron energy loss spectroscopy)  experiments with the same spectrometer at 30 keV and 200 eV pass
energy where the obtained resolution is 0.5 eV FWHM \citep{Werner2007}.  The poorer energy resolution  is due to both the
larger size of the emitting surface (4 mm for this experiment, compared to  0.2 mm of the
electron beam in REELS) and the fact that ripple and drift in the main high-voltage (HV) power supply cancels out in
 REELS  ( incoming electrons are accelerated, outgoing electrons are decelerated by the main HV potential) but not in  the experiment described here.

The generally accepted explanation for the tail at the low energy side of the main peak is
electronic excitations either created as part of the excitation process itself (referred to as `shake' 
electrons in atomic physics or `intrinsic plasmon losses' in condensed matter  vocabulary) or created during
transport of the electron out of the sample (e.g. creation of (surface-)plasmons). As the iodine atom is absorbed right
at the surface, the contribution of bulk plasmons should be small, and the probability of
surface plasmon creation at these high energies is of the order of 3\% \citep{Chen2002}. Consistent with this, the line shape is not very sensitive for the experimental geometry, as rotating the sample over 60$^\circ$ changes the spectrum only slightly (Fig. \ref{fig:l1-high-res}, top panel). Thus it is expected that shake
effects  are the main cause of the tail at the low-energy side. Using
the parameters of Table \ref{table:fit}, one would conclude that the sum of the areas of the tail
components is almost equal to that of the main component.  There is significant overlap
of the tail components and the main component and other fitting approaches could produce a 
different area of the tail component.  It is clear, however, that the area of the tail is at least half the area of
the main component. In comparison, the intrinsic intensity for high-energy photoemission of carbon was found to be 58\% of the total intensity \citep{Kunz2009} very similar to the value found here.
In the context of medical physics, these `shake' electrons are important as they
could provide a significant source of additional  low-energy electrons with their  large genotoxic effects 
\citep{Boudaiffa2000}.

\begin{table}
\begin{tabular}{|c|c|c|}
	\hline 
		& exp. & calc. \\ 
		\hline
	L$_1$:L$_2$:L$_3$ & 1:0.085(3):0.019(3) & 1:0.082:0.025 \\ 
	L$_1$:M$_1$ & 1:0.202(5) & 1:0.198 \\ 
 	M$_1$:M$_2$:M$_3$ & 1:0.095(6):0.023(7) & 1:0.087:0.026 \\ 
	M$_1$:N$_1$ & 1:0.18(2) & 1:0.199 \\ 
	L$_1$:KL$_2$L$_3$ & 1:0.60(1) & 1:0.526 \\ 
	\hline 

	\hline 
\end{tabular} 
\caption{The intensity ratio of the various CE and Auger components compared with those given in the literature and  as calculated based on BrIcc \citep{Kibedi2008} and BrIccEmis \citep{Lee2016} using mixing ratio $\delta=0.015$ and  penetration parameter $\lambda=-1.2$, the evaluated nuclear parameters  based on these and literature measurements.   }
	\label{table:results}
\end{table}

In order to measure the L$_1$, L$_2$ and L$_3$-CEs in a single measurement, the spectrometer was operated  in the low-energy resolution mode.  The result is shown in Fig. \ref{fig:l1-l3-low-res}. All three conversion lines can be easily
identified. The L$_2$ and  L$_3$ lines are considerably weaker than the L$_1$ line.
The intensity ratios of the various CE contributions reflect
the initial vacancy population of the Auger cascade after decay of the excited state of $^{125}$Te. The L$_1$ part of the spectrum was analyzed
using the same parameters for the peak shape as in Table \ref{table:fit}, but the experimental
resolution was a free fitting parameter.  An experimental resolution of 6.6 eV was obtained in this way. The L$_2$ and L$_3$ CE peaks can then be fitted with  only two additional  fitting parameters (their intensities), as their relative  position  follows directly from the splitting of the L$_1$, L$_2$ and L$_3$ levels and their lifetime broadening is taken from the literature.  The resulting intensity ratio is shown in Table \ref{table:results}.

\begin{figure}
	\centering
	\includegraphics[width=7.5cm]{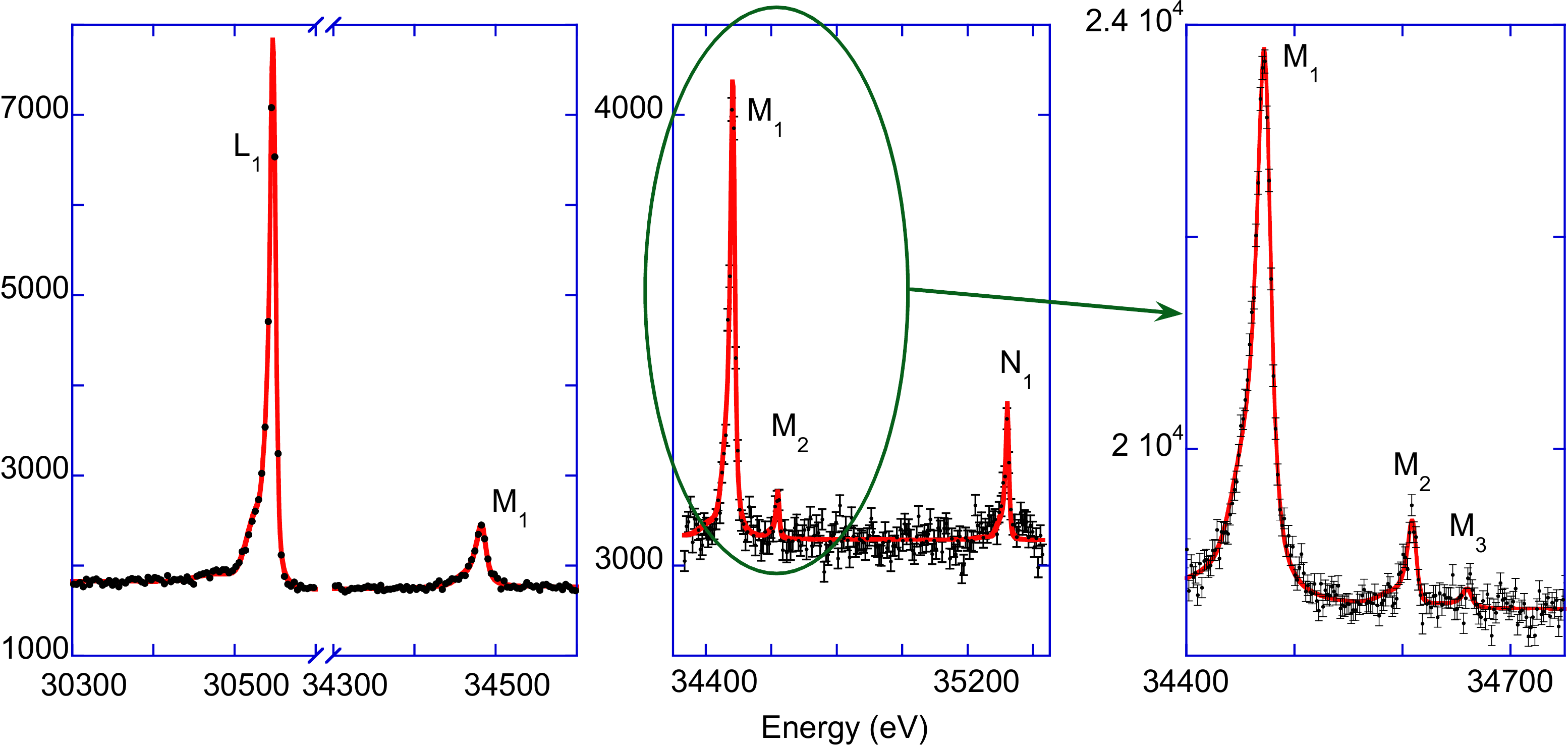}
	\caption{The left panel shows the spectrum of the L$_1$ and M$_1$-CE lines.  The central panel is a measurement that shows the M$_1$, M$_2$ and N$_1$-CE lines.  The right panel is run with better statistics for the M conversion electrons only.}
	\label{fig:m1fig}
\end{figure}

The L$_1$ to M$_1$ intensity ratio was determined in a different measurement, as shown in the left panel of Fig.
\ref{fig:m1fig}.  As the peaks are almost 4 keV apart, a 3 keV region in between the
peaks was skipped to reduce the measurement time.  Again the M$_1$ peak was fitted with the same set
of Gaussians as the L$_1$ peak, but with a larger Lorentzian broadening (10.2 eV), as the M$_1$ lifetime is much shorter. The results are also listed in Table
\ref{table:results}.

The intensity ratio of the M$_1$, M$_2$ and N$_1$ conversion lines was established in the same way (Fig. \ref{fig:m1fig}, center and right panel).    The M$_3$ intensity was very low and the determination of its area has a 30\% statistical error.  

The intensity ratios can be used to extract nuclear structure parameters $\delta$ and $\lambda$. This is discussed in a separate paper \cite{Tee2019}, but the calculated intensity ratios, based on nuclear parameters that result in the `best' fit of the present and literature data are shown in table \ref{table:results} as well.  Based on these nuclear structure parameters one can then calculate the number of the various conversion electrons per nuclear decay as well as the fraction of the Te excited state that decays by the gamma rays emission (we obtain 6.83\%, slightly larger than the previously evaluated literature value  (6.68\%) ).  We will use the L$_1$ and K conversion line intensities  to fix the intensity scale of the Auger electrons.

\begin{figure}
	\centering
	\includegraphics[width=1.0\linewidth]{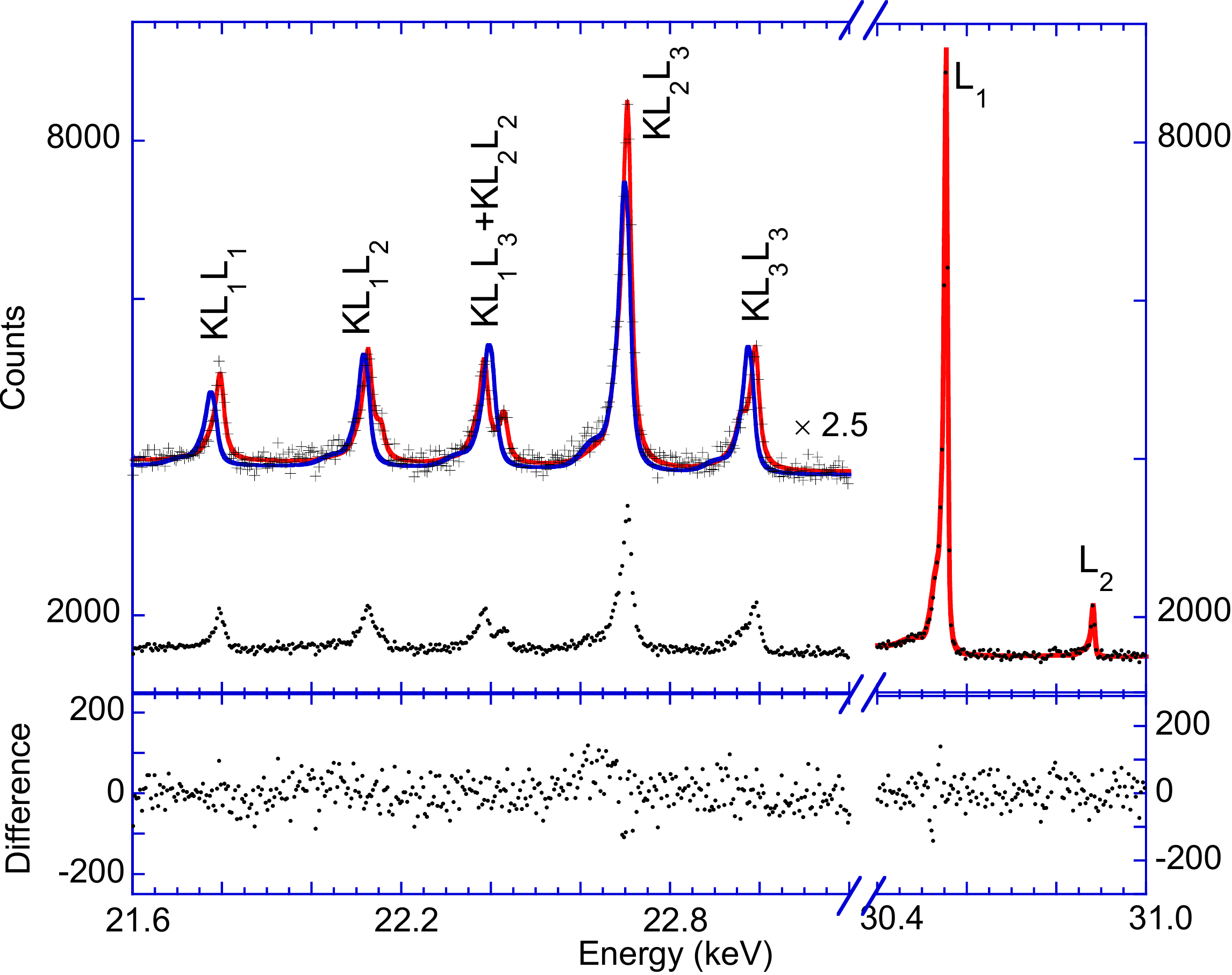}
	\caption{The KLL Auger and L$_1$, L$_2$-CE spectrum as measured in a single run.  The red line shows a fit of the KLL spectrum with 8 peaks. All peaks have the same tail parameters as in Table \ref{table:fit}, the residual is shown in the lower panel. The blue line shows a description of the KLL spectrum based on the BrIccEmis calculation which was scaled such that the L$_1$ part of the  spectrum has the same area at the experiment. }
	\label{fig:klll2fig}
\end{figure}

\begin{table*}[]
	\footnotesize
	\centering
	\scriptsize
	\begin{tabular}{ccccccc}
		\toprule
		\multirow{2}{*}{Transition}                                                  & \multicolumn{3}{c}{Rel. energy to the measured $KL_2L_3(^1D_2)$ (eV)}                         & \multicolumn{3}{c}{Relative Intensities}             \\ \cmidrule(l){2-4} \cmidrule(l){5-7}
		& Exp        & BrIccEmis               & Ref. \cite{Larkins1977} & Exp       & BrIccEmis              & Ref. \cite{Chen1980} \\ \cline{1-7}
		\hspace*{0pt}$KL_1L_1(^1S_0)$                                                             & -902  & -834                  & -892            & 0.262(5)  & 0.263                  & 0.263           \\ \hline
		\hspace*{0pt}$KL_1L_2(^1P_1)$                                                             & -574  & \multirow{2}{*}{{\large \}} -492} & -571            & 0.296(10) & \multirow{2}{*}{{\large \}} 0.397} & 0.305           \\
		\hspace*{0pt}$KL_1L_2(^3P_0)$                                                             & -551  &                         & -537            & 0.086(6)  &                        & 0.093           \\ \hline
		\hspace*{0pt}$KL_1L_3(^3P_1)$                                                             & -312  & \multirow{2}{*}{{\large\}} -212} & -314            & 0.309(7)  & \multirow{2}{*}{{\large \}} 0.457} & 0.289           \\
		\begin{tabular}[c]{@{}c@{}}\hspace*{-0pt}$KL_1L_3(^3P_2)$+\\ \hspace*{-0pt}$KL_2L_2(^1S_0)$\end{tabular} & -279  &                         & -265            & 0.153(6)  &                        & 0.168           \\ \hline
		\hspace*{-0pt}$KL_2L_3(^1D_2)$                                                             & 0  & 90                  & 6            & 1         & 1                      & 1               \\ \hline
		\hspace*{-0pt} $KL_3L_3(^3P_0)$                                                             & 246 & \multirow{2}{*}{{\large \}} 366} & 253            & 0.071(6)  & \multirow{2}{*}{{\large \}} 0.436} & 0.077           \\
		\hspace*{-0pt}$KL_3L_3(^3P_2)$                                                             & 293  &                         & 287            & 0.364(7)  &                        & 0.360           \\ \toprule
	\end{tabular}
	\caption{Energies and relative intensities of the $KLL$ Auger transitions from $^{125}$I electron capture decay.}
	\label{table:kll}
\end{table*}
\subsection{KLL  Auger Electrons}
We now focus on the KLL Auger electrons and their intensities relative to the L$_1$-CE line. The relevant spectrum  is shown in Fig. \ref{fig:klll2fig}.  These data are similar to those
obtained with a  magnetic spectrometer by \cite{Graham1962}.  The KLL Auger spectrum consists of a
number of peaks spread over more than 1 keV. There are (at least) two ways of describing these
spectra:  \\
(i) One can characterize each final state in terms of the atomic orbitals they originate from and to the total
angular momentum and total spin quantum number of the final state  \cite{Larkins1977}.  This leads to 9 possible final
states in the intermediate coupling scheme: $^1$S$_0$ (K-L$_1$L$_1$), $^3$P$_2$ (K-L$_1$L$_3$),
$^3$P$_1$ (K-L$_1$L$_3$), $^1$P$_1$ (K-L$_1$L$_2$), $^3$P$_0$ (K-L$_1$L$_2$),
$^3$P$_0$ (K-L$_3$L$_3$), $^1$S$_0$ (K-L$_2$L$_2$),  $^3$P$_2$ (K-L$_3$L$_3$),
$^1$D$_2$ (K-L$_2$L$_3$).  This approach was followed by \cite{Larkins1977}.
The energy of the  final-state energy can be deduced from the energy of two single L holes plus their correlation energy,
which depends on the total L and S quantum numbers and is calculated by Slater integrals.  This
approach works well for two core holes, but becomes  cumbersome when more vacancies are present, later
in the cascade.   \\
(ii) One can neglect the fine splitting   and characterize the final state
in terms of L$_x$ only.  Then there are 6  possible final states 
(L$_1$L$_1$, L$_1$L$_2$, L$_1$L$_3$, 
L$_2$L$_2$, L$_2$L$_3$, L$_3$L$_3$) but for Te the 	L$_1$L$_3$ and L$_2$L$_2$ energies are almost
identical and these two contributions will not be resolved. The spectrum  consists then  of 5
peaks.  This approach is adopted in  BrIccEmis \citep{Lee2016} and remains manageable
when one calculates  several steps down in the relaxation cascade, when more vacancies are present.

A good fit of the KLL Auger spectrum required  8  components (see Fig. \ref{fig:klll2fig}).  The Auger lifetime
broadening, i.e. the sum of the lifetime broadening of the K level and two L levels,  was taken from ref.
 \cite{Krause1979a}, and the peak shape of Table \ref{table:fit} was used (assuming
thus that shake effects are the same for a CE and an Auger electrons).   There is some excess intensity in the measurement
visible $\approx$ 70 eV below the main Auger line, and the fit quality would improve somewhat if  the
intensity of tail component 3 of Table \ref{table:fit} is doubled.

The extracted parameters of the fit are reproduced in table \ref{table:kll}. When comparing with BrIccEmis one has to sum certain peaks. Overall the intensities are well-reproduced and there is good agreement with both the theory of Chen {\it et al.} \cite{Chen1980} and BrIccEmis \cite{Lee2016}. The energy deviates by $\approx 10$ eV from the semi-empirical estimate of Larkins \cite{Larkins1977}, but by $\approx 90$ eV with the one calculated from in a Dirac-Fock approach by BrIccEmis mainly due to the fact that the Breit-interaction and QED corrections were not implemented in the calculation. The splitting of the multiplets as given by Larkins agrees quite well with the experiment, it is only $\approx 1$\% smaller than the observed ones. The BrIccEmis calculated splitting (relative to the main KL$_2$L$_3$ component) is also quite good, it  is   too large for the KL$_1$L$_1$ and KL$_1$L$_2$ by slightly more than 1\%, and for the other components the differences are small and the trend is less clear. 

In the fit the experimental  energy
resolution  was taken  as a fitting parameter and  one obtains slightly different values for the resolution of the
L$_1$ peak and the KLL Auger peaks (6 eV and 10 eV, respectively).  There is, however, one additional
cause of broadening that has not been taken into account here, namely that the KLL Auger spectrum consists of two,
slightly different contributions of almost equal intensity:\\
(i) Auger decay of a K hole  due to electron capture by $^{125}$I, which  results in the formation of a Te nucleus.  Here the valence electrons may not have adjusted to the new
nuclear charge, resulting in a slightly different electronic structure (and hence KLL Auger energy)
than for Te in the ground state.\\
(ii) Auger decay after a K hole is created when the  $^{125}$Te nucleus decays to its ground state by
internal conversion. 

The energy difference between these two Auger decays is of the order of 10 eV \citep{Kovalik1998}, and
this could be the cause of the observed additional broadening. If the energy resolution is fixed at
the value derived from the L$_1$ feature taken in the same run (6 eV), and one fits the Auger part with two
equal contributions, shifted slightly in energy, then the best fit is  for a shift of $7 \pm 1$ eV, a value
surprisingly close to what was obtained by Kovalik {\it et al.} \citep{Kovalik1998} for Xe.

The BrIccEmis program provides a complete description of the Auger spectrum.  In Fig. \ref{fig:klll2fig} the results of the BrIccEmis
calculation are  shifted  down by 90 eV. The vertical scale for the 
BrIccEmis calculation was chosen to fit  the L$_1$-CE peak height of the experiment. Besides the
absence of some of the fine-structure in this calculation, it is again clear that BrIccEmis, using transition rates from EADL \citep{Perkins1991}, somewhat underestimates the Auger intensity relative to the L$_1$-CE line.

The observed intensity ratio  of the L$_1$-CE line to the main component of the KLL Auger spectrum
($KL_2L_3(^1D_2)$)  is  1:0.61(1).  The value calculated from BrIccEmis is 1:0.53, and the calculated Auger
intensity is thus slightly lower than the experimentally observed one.  The peak-area ratio obtained in this way is reproduced in Table \ref{table:results} as well.  As discussed elsewhere \cite{Alotiby2018} this intensity ratio is very sensitive to the K-shell fluorescence rate assumed. A slight decrease of this quantity (within its experimental uncertainty) from 0.878 to 0.85 resolves this issue.

\begin{figure*}
	\centering
	\includegraphics[width=14cm]{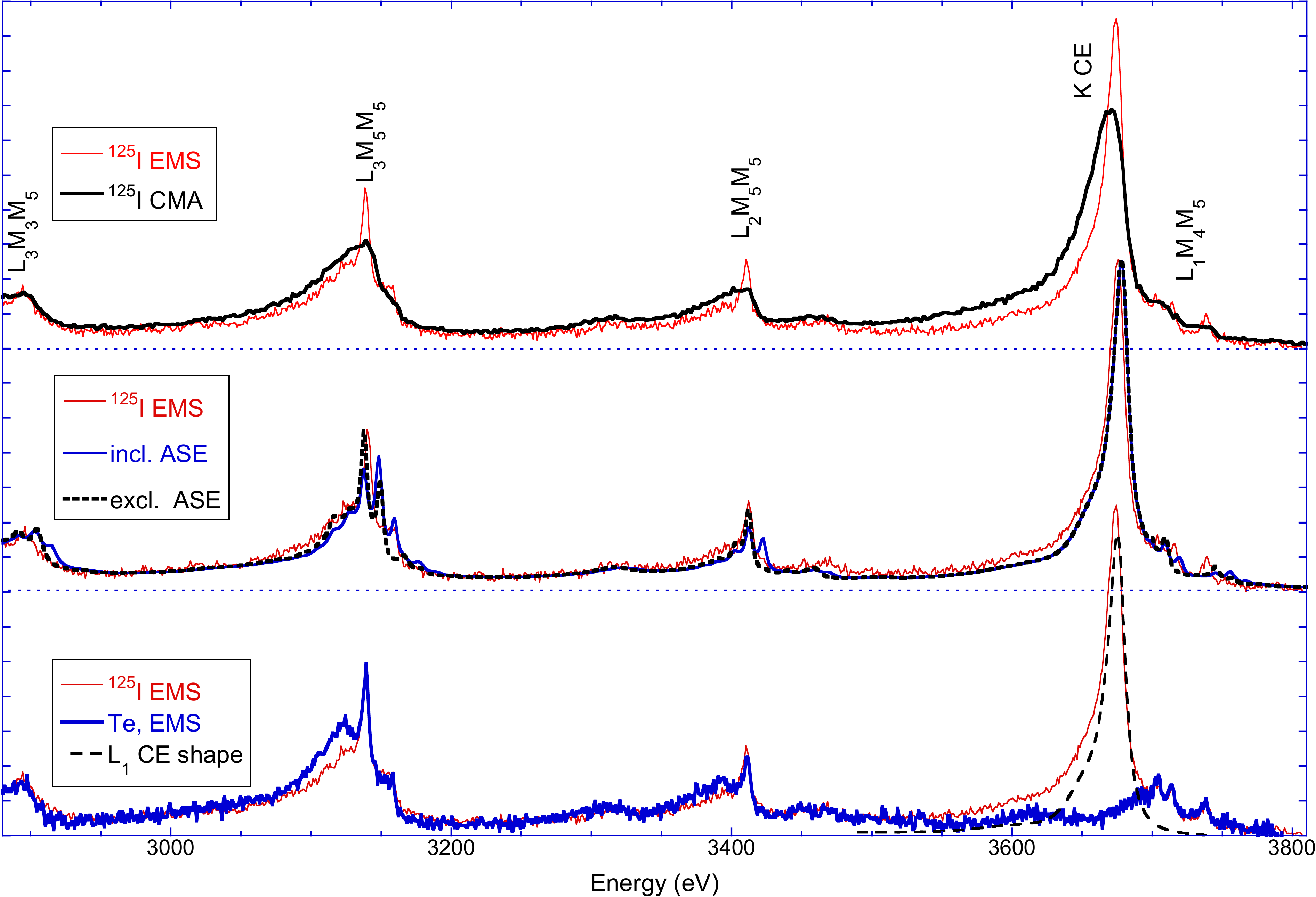}
	\caption{The top panel shows the Te LMM Auger spectra induced by the decay of $^{125}$I LMM Auger spectrum as measured with the CMA (thick line) and compared with the spectrum measured with the EMS spectrometer (thin lines).  Some of the major transitions are indicated.  The central panel compares the EMS measurement with BrIccEmis simulations with and without atomic structure effect (ASE).   The lower panel compares the Te LMM Auger spectra induced by nuclear decay, with the electron-beam induced one. In the latter the conversion line is absent. The dashed line is the shape of the L$_1$ CE line, for comparison with the K CE line. The L$_1$ CE shape has been adjusted to take into account the difference between the L$_1$ lifetime (2.2 eV) and K lifetime  (9.9 eV). }
	\label{fig:lmmcomp-crop}
\end{figure*}

\subsection {K conversion  and LMM Auger spectra} 
The K CEs have an energy of 3.679 keV and are in the same energy range as the Auger electrons originating from vacancies in the L shell. This energy range can be measured with both spectrometers.  In Fig. \ref{fig:lmmcomp-crop} top panel, we compare spectra measured with both spectrometers.  The EMS spectra were measured in the high-energy resolution mode and hence are restricted to cover a range smaller than 1 keV.  Clearly the energy resolution of the EMS spectrometer is superior to the CMA.  The pass energy of the CMA was 100 eV in this measurement. Lowering the pass energy appears at first sight a way to increase the energy resolution, but in practice caused increased intensity at the low-energy side of the peaks. The cause of this is probably the following. Upon entering the second stage in the CMA the electrons are decelerated by a field in between  two spherical grids with slightly different radius. The energy of the electrons is thus reduced very severely (by a factor of 35 for 100 eV pass energy, twice as much for 50 eV pass energy).  Under these conditions aberrations due to micro lenses formed by the grid will become important \cite{Read1999} and are suspected to be the cause of the increased intensity at lower energy of the main peak.

The EMS spectrum is used for a detailed comparison with theory. BrIccEmis simulations were done to calculate the intensity of the Auger and K CE lines. A spectrum was generated assuming that the tail was the same as observed in the L$_1$ CE spectrum.  The lifetime broadening of each peak was calculated from the ref. \cite{Perkins1991}. The resulting simulated spectrum  is shown in the central panel.  

The simulation was done under two assumptions: The LMM Auger after electron capture and internal conversion both occur in an atom with the Te atomic structure (no atomic structure effect) or in a Te atom for the conversion electron or a I atom after electron capture (including atomic structure effect).  Clearly the latter simulation produces more peaks, but neither describes the spectrum perfectly, as is expected for the simplified coupling scheme used in BrIccEmis. Overall the intensity distribution is described quite well in the simulation, in particular the K-CE to Auger peak intensity ratios but in between the peaks the intensity in the measurements exceeds that of the simulation slightly.

\begin{table}
	\center
	\begin{tabular}{|c|c|c|}
		\hline
		~               & ${\rm L}_2/{\rm L}_1$ & ${\rm L}_3/{\rm L}_1$ \\ \hline
		$^{125}$I  & 1.43                                            & 2.60                                            \\
		29 keV e$^-$ & 1.40                                            & 3.14                                            \\ \hline
	\end{tabular}
\caption{The  population of the L$_2$  and L$_3$ level relative to the population of the L$_1$ level after decay of $^{125}$I (sum of both cascades, calculated using the same data bases  as BrIccEmis) and as induced by 29 keV electrons, as calculated using the program from Bote \cite{Bote2009a}. } \label{table_Lpop}
\end{table}
It is insightful to compare the $^{125}$I induced spectra  with those obtained by electron beam from Te films.  The initial L vacancy distribution for both radioactive decay and electron bombardment is given in table \ref{table_Lpop}.  Note that the initial population is given here, redistribution within the L shell due to the Coster-Kronig process is not considered.  The fact that the relative population in both cases are so similar is purely accidental as the processes that lead to the vacancy production are completely different. For example,  the L$_1$ population in the case of $^{125}$I decay is mainly due to L$_1$ CE emission and IC, whereas the population of the L$_2$ and L$_3$ shell is mainly  due to radiative  decay of the K vacancies.  Neither of these processes occur for the electron beam case.

Indeed both the electron beam and the $^{125}$I derived Auger spectrum look very similar with the exception of  the K CE  peak present that is only present in the latter case (Fig. \ref{fig:lmmcomp-crop} lower panel).  The main difference is that there is additional intensity near 17 eV below the main Auger lines. As the evaporated Te film has a thickness of the order of the inelastic mean free path of the LMM Augers the probability of extrinsic plasmon excitations is now significant causing the excess intensity shifted by the Te plasmon energy (17 eV). By operating the spectrometer in the electron-energy loss mode \cite{went2006} (analyser tuned to energies near the gun energy) we find that these samples show indeed a plasmon energy near 17 eV, in agreement with literature REELS data for Te \cite{Werner2009}.

 For comparison also the calculated shape of the K conversion line was plotted based on the parameters of table \ref{table:fit} adjusted for the lifetime of the K level.  It follows indeed closely the difference of the electron beam induced and $^{125}$I derived case.

For the electron-beam induced Auger spectrum of the  Te film there is by definition no atomic structure effect. In the case of $^{125}$I decay the theory predict that, if the valence electrons do not rearrange themselves at the timescale of the LMM Auger decay, its shape should be affected quite noticeable by the atomic structure effect.  The fact that the $^{125}$I and the Te electron-beam induced Auger spectra are so similar means that the valence band relaxation is faster than the LMM Auger process and the atomic structure effect plays no significant role for the $^{125}$I induced LMM spectra.


%
\begin{figure}
	\centering
	\includegraphics[width=7.5cm]{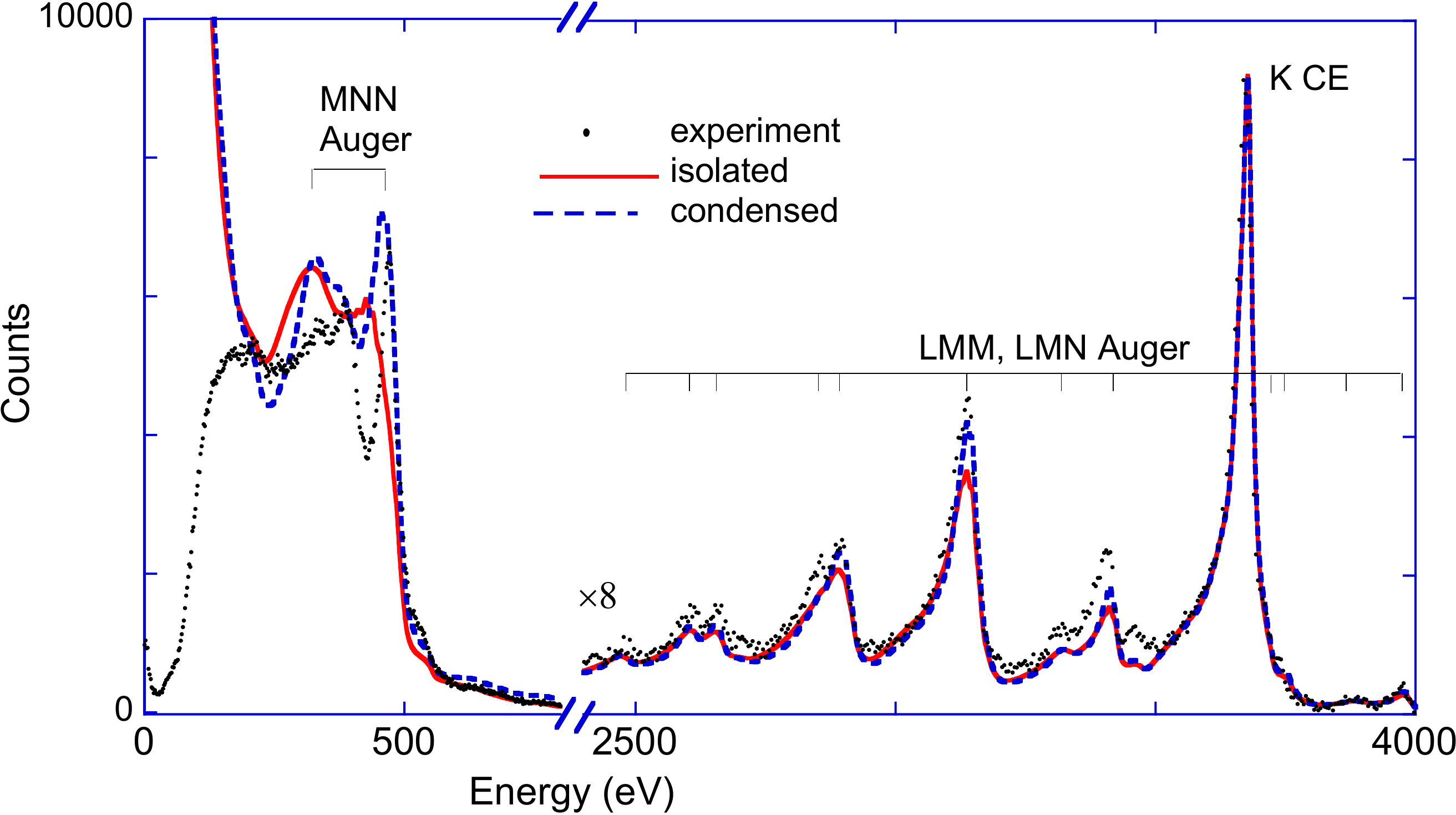}
	\caption{A comparison of the measured spectra compared with theory assuming either 'isolated' $^{125}$I (where the charge keeps accumulating) or condensed $^{125}$I (where vacancies in the valence band are assumed to be reoccupied immediately).  For the comparison with theory an energy dependence of the spectrometer's efficiency was assumed to scale as $1/E^{0.8}$.}
	\label{fig:isolatedcondensed}
\end{figure}

\begin{figure}
	\centering
	\includegraphics[width=7cm]{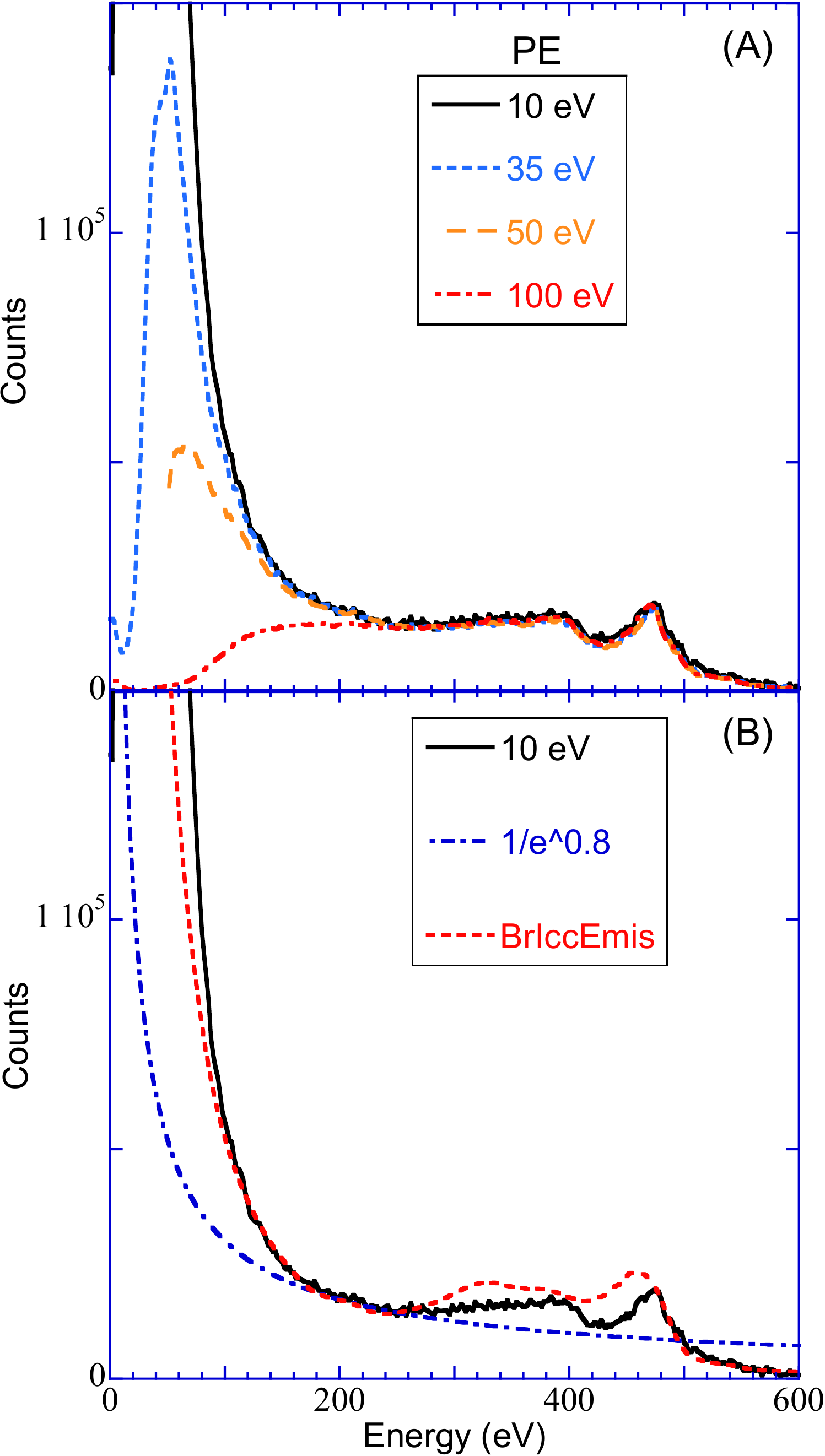}
	\caption{In (A) we show the effect of the pass energy (in eV, as indicated) on the spectrum at low energies. All spectra are normalised to equal height near 470 eV. In (B) we show the result 10 eV PE results together with the    BrIccEmis calculation for the condensed phase scaled in the same way as in Fig. \ref{fig:isolatedcondensed}. The dash-dotted line, proportional to $1/E^{0.8}$  indicates how the transmission of the analyser changes with energy, the BrIccEmis results were scaled by this factor.}
	\label{fig:passenergy}
\end{figure}

\subsection{MNN and NXX Auger intensities}
The CMA spectrometer allows us to measure a complete spectrum from 0 to 4 keV. In particular we can use it to compare the LMM and MNN intensities. (Below 1 keV there are many overlapping Auger series contributing to the intensity. We refer to the intensity by the main contribution, see e.g. ref. \cite{Lee2016}. ) This is done in Fig. \ref{fig:isolatedcondensed}.  The theory from BrIccEmis was scaled by $1/E^{0.8}$ which should take into account the energy dependence of the analyser transmission above 200 eV.  Experiment and theory were again scaled by the K CE line.  However, now we had to significantly adjust the tail  parameters in order to get a good description of the K-CE and LMM Auger peak shapes observed in the  experiment, and even then the level of agreement with the Auger intensity was not as good as in the EMS case, in particular the measured Auger intensity appears 15-20\% larger than the calculated one.  We take as an indication that either the tail description or the Shirley-type background assumed needs more refinement. 

The BrIccEmis program was run using two different assumptions:\\
-When vacancies propagating towards the outer shell  they are filled instantaneously when they reach the  valence band. This is called the `condensed' model as it is assumed to be valid in the condensed phase.\\
-Vacancies accumulate in the outer shell and a multiple charged ion is formed in the final state with a charge state corresponding to the total number of electrons (Auger and CE) emitted. This is called the `isolated' model and is thought to apply to the gas-phase.  \\For the LMM part of the spectrum the difference is slight, the `condensed' model has slightly more intense  peaks, and is in slightly better agreement with the experiment.  
Around 500 eV the measured intensity increases sharply, as here the MNN Auger start contributing.  Now the condensed and isolated model calculations differ more substantially.  The condensed model has a clear peak near 500 eV, whereas the isolated model has a shoulder at somewhat lower energies. The experiment shows a clear peak at slightly over 500 eV, and is thus much closer to the condensed model than the isolated one. Clearly the agreement is not as good as for the KLL and LMM Auger spectra, reflecting the fact that now we are further down the cascade and the interaction with the many vacancies present is accounted for only approximately.

At energies below 200 eV the experiment with 100 eV pass energy start deviating strongly from the theory.  This is expected as then the transmission of the CMA drops sharply (see Fig. \ref{CMA_Simion_results}, lower panel).  By changing the pass energy one can change the energy where the transmission  of the CMA  does not scale as $1/E^{0.8}$ anymore but drops off sharply.  This is demonstrated in Fig. \ref{fig:passenergy}.  The measurements at different pass energies  were normalised to equal height of the peak at 470 eV. There is no obvious sharpening of this peak with decreasing pass energy, indicating that most of the observed width is intrinsic. For lower pass energies a region of large intensity develops at low energies.  Interpretation here is not so simple.  Part of it is due to  $1/E^{0.8}$ dependence of the analyser transmission but this only explains a part of the strong enhancement seen at low energies. The excess intensity could either be interpreted as very low energy Auger electrons, or in terms of secondary electrons emerging from the Au film, generated by the interaction of  energetic Auger electrons with the Au valence electrons. The BrIccEmis simulation predicts also sharp increase near 100 eV and it that case the steep increase is  due to Auger electrons with the initial vacancy in the N shell. The agreement with BrIccEmis is quite good for the low pass energy measurements. From this one would conclude that Auger electrons dominate the intensity at least down to 50 eV.

\begin{figure}
	\centering
	\includegraphics[width=7.5cm]{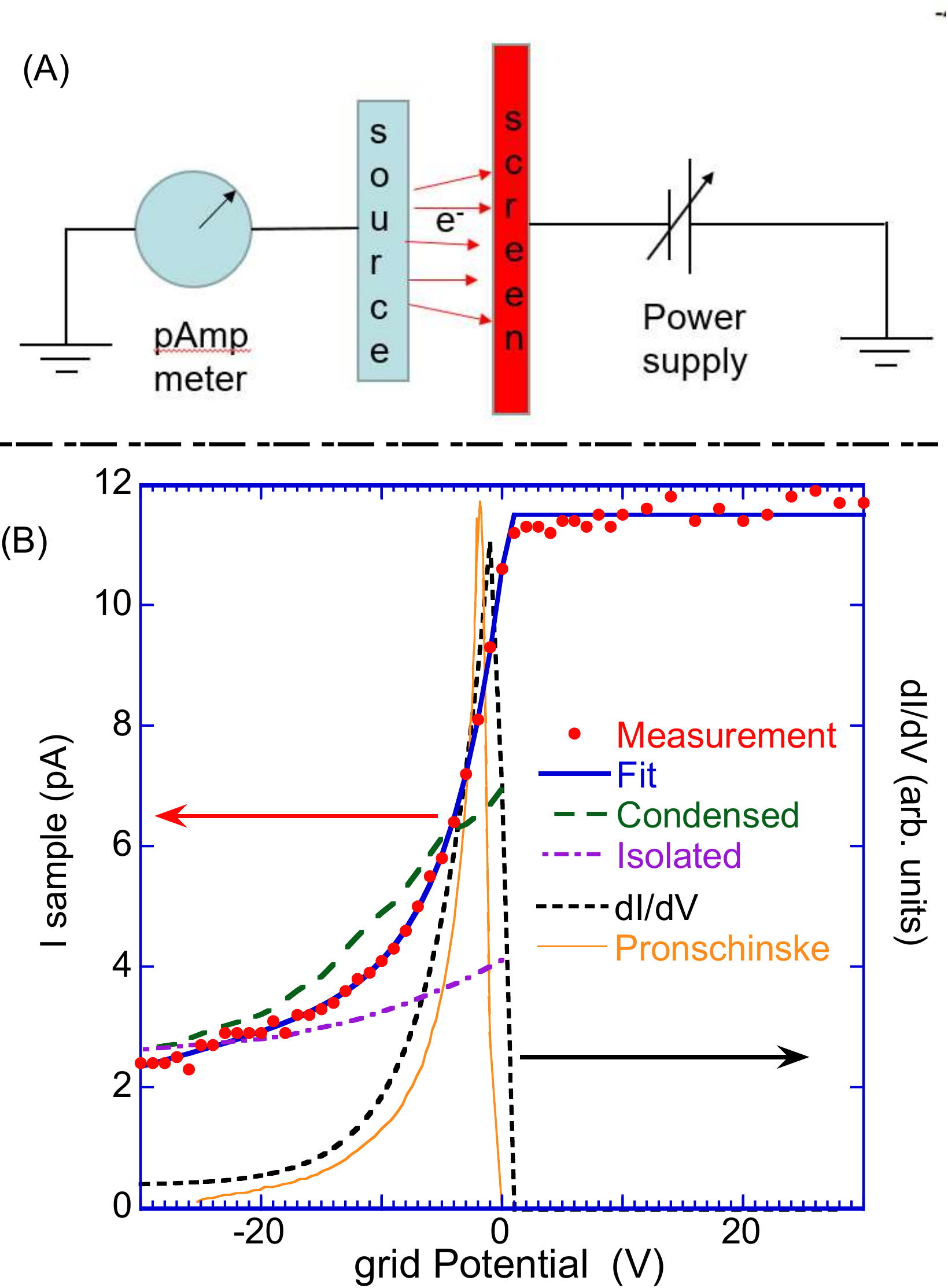}
	\caption{In (A) we show schematically how the current leaving the sample was measured, as a function of the voltage applied to a high-transparency  screen. In (B) we show the measured current (dots) fitted by a semi-empirical function. The derivative of this function (dashed line) corresponds to the energy distribution of the low-energy electrons leaving the sample. The thin yellow line is the low energy tail of the spectrum as measured by Pronchinske{\it et al.} \cite{Pronschinske2015}}
	\label{fig:currentmeas-crop}
\end{figure}

\subsection{Differential current measurements at low energies}
The total number of Auger electrons emitted per decay  according to BrIccEmis is 20 electrons using the  condensed phase approximation and 11.9   electrons using the isolated atom approximation (plus  0.93 CE per decay) \cite{Lee2016}.  Half of these electrons are expected to move into the Au film.  For a 4.4 MBq source (the activity  at the time of this measurement) this would correspond, using the `condensed approximation' to a current of $ 20/2\times 4.4\times 1.6~10^{-13}\approx7$ pA. The observed current was 60\% larger (11 pA). A sharp decline of the current with the applied voltage was observed but this does not necessarily imply that the current is dominated by secondary electrons.  The average kinetic energy of NXX Auger electrons, as simulated by BrIccEmis in the condensed phase is 14 eV \cite{Lee2016}.  As the NXX Auger spectrum extends up to 100 eV, many NXX Auger   electrons will have energies far less than 14 eV.

The current as predicted by the BrIccEmis simulation  was calculated assuming  that half the produced Auger electrons have an initial velocity directed towards the Au film, and will thus not contribute to the current and that when a voltage $-V_{\rm repel}$ is applied to the screen only electrons with kinetic energy larger than $V_{\rm repel}$ will escape.  The agreement near -25V is surprisingly good, and at lower magnitude suppression voltage the BrIccEmis are lower than the measured current, for both the condensed and isolated approximation.  This would mean that about half of the total current leaving the sample is due to secondary electrons.

Thus the most straight-forward interpretation is that most of the current observed is due to Auger emission and the enhancement due to the Au substrate is much less pronounced than found by Pronschinske {\it et al.} \cite{Pronschinske2015} who concluded that the Au surface increased the low-energy flux six-fold.

Note that our experience with this type of measurements is limited, and fully quantitative interpretation could be affected by unknown errors (e.g. secondary electrons from the grid should affect the measurement at some level, even when one uses a highly open, carbon coated mesh as was done here, we measure in reality $p_\perp^2/2m_e$ not $p^2/2m_e$ as assumed in the analysis, influence of the work function of the surface).  However, the level of agreement in these  somewhat preliminary measurements is promising.

\section{Summary} 
The measurement described here have implications to our understanding of various aspects of the Auger decay. We summarise them here succinctly:\\

\noindent Line shapes\\
For the analysis of the  ratio of the areas of the Auger and CE lines, the assumed line shape is
crucial.  Clearly, the observed peaks deviate strongly from a pure Voigt line shape, a fact also
evident from other Auger measurements (e.g. \cite{Kovalik1998}) and photoemission measurements of deep
core holes (\cite{Piancastelli2017}). Here the approach generally was to minimize the number of
fitting parameters by assuming that the Gaussian part of the line shape of the L$_1$-CE peak applies
to all features, and only the Lorentzian width was  varied according to the level lifetime.  For
the KLL Auger part there are some indications that the tail intensity is somewhat larger.

Using the EMS spectrometer the same approach worked quite well for the LMM Auger spectrum.  Here we could obtain an Auger spectrum both after decay from $^{125}$I and electron-beam induced Auger from a thin (40 \AA\ thick) Te film. The main difference was the absence of the K conversion line.  The shape of this difference had the same energy distribution as the L$_1$ lineshape adjusted for the different lifetime. \\ 

\noindent L$_1$ CE -  KLL Auger electron intensity ratio\\
The combined  L$_1$ CE - KLL Auger
measurement  indicate that in the experiment  the relative Auger intensity is about 15-20\% too high compared
to the calculated one.  This has been attributed to  a small error in the fluorescence yield used in BrIccEmis \cite{Alotiby2018}.\\

\noindent K CE - LMM Auger electron intensity ratio\\
When measured with the EMS spectrometer we can analyse the K CE to LMM intensity using the same tail description as the L$_1$ CE peak.  Then a good agreement is found for the intensity of the Auger and K CE electrons.  When using the CMA the peak shape is different, and adjusting the tail description so the right shape is obtained we find that the calculated  Auger intensity is about 15\% less than the observed one. This is probably an indication that the peak shape used is not 100\% correct.\\

\noindent K CE - MNN and  -NXX Auger electron intensity ratio\\
Here the comparison depends critically on the dependence of the analyser transmission on energy. Using a $1/E^{0.8}$ dependence, as obtained from SIMION simulation we get agreement  on a 20\% level, although the energy position of the main peak is 20 eV off.  For the NXX intensity we have to extrapolate using lower pass-energy measurements. The measurement at 10 eV pass energy increases more quickly with decreasing energy than the calculated one. This could  indicate that some of the intensity seen here could be better described as secondary electrons leaving the Au film.\\

\noindent Current measurement \\
Based on the current measurement to the sample, and its dependence on the voltage on a closely located grit, it appears that the current leaving the sample is about 50\% larger than calculated using BrIccEmis. This can be attributed to secondary electrons leaving the Au film. We do not think that the Au surface increases the flux of emitted electrons 6-fold, as suggested by  Pronschinske {\it et al.} \cite{Pronschinske2015}.\\

\noindent Isolated versus condensed\\
The onset and the shape of the MNN Auger spectrum is much closer to that simulated within the condensed approximation than the isolated approximation. Build-up of multiple vacancies in the outer shell during the cascade (as is the case in the isolated model) would cause a shift of the MNN Auger to lower energies. Thus the `condensed model' describes the experiment better.\\

\noindent Atomic structure effect\\  
For the KLL Auger the fit improved if one assumes that the Auger lines after electron capture differ by 10 eV in kinetic energy from those emitted after conversion electron emission.  As 7 eV is less than the life-time broadening of the KLL Auger this shows up as an additional broadening of the Auger spectrum, that, in-principle, also could be explained by assuming that the energy resolution for Auger would be not as good as for the L CE measurements. 

Some of the LMM lines are sharper than 7 eV.  Here we see no evidence of an atomic structure effect and their shape for $^{125}$I  and for electron-beam induced emission of Te films  can be understood based on energy loss in the 40~\AA\ thick Te films.  Note that the LMM Auger happens at a longer time scale than the KLL Auger.  Therefore the presence of the atomic structure effect for the KLL Auger and the absence of the atomic structure effect for the LMM Auger is not necessarily a contradiction.  

\section{Conclusion}
We found that the BrIccEmis program gives generally a good description of the observed Auger spectra over the full energy range. Especially when using the CMA at lower energies our conclusion depends quite strongly on the analyser transmission used, but using the transmission  derived from SIMION simulations we obtained good agreement.  No indication of very large enhancement of the low energy flux due to the Au surface was found.

\section{Acknowledgments}
This research was made possible by  Discovery Grant DP140103317 of the Australian Research Council.  The authors want to thank Mr Ross Tranter  and Mr Justin Heighway for help preparing the mesh for the low energy electron current measurement experiment.

\end{document}